%% file: paper.tex
\documentclass[10pt,conference]{IEEEtran}
\IEEEoverridecommandlockouts
\usepackage{cite}
\usepackage{amsmath,amssymb,amsfonts}
\usepackage{algorithmic}
\usepackage{graphicx}
\usepackage{textcomp}
\usepackage{xcolor}

\usepackage{tabularx}
\usepackage{booktabs}
\usepackage{xspace}
\usepackage{url}

\def\BibTeX{{\rm B\kern-.05em{\sc i\kern-.025em b}\kern-.08em
    T\kern-.1667em\lower.7ex\hbox{E}\kern-.125emX}}
    
\IEEEoverridecommandlockouts\IEEEpubid{\makebox[\columnwidth]{ 978-1-6654-3540-6/22~\copyright~2022 IEEE \hfill} \hspace{\columnsep}\makebox[\columnwidth]{ }}
\begin{document}

\title{Zero-Rating, One Big Mess: Analyzing Differential Pricing Practices of European MNOs
\thanks{This work was funded through the NGI0 PET Fund, a fund established by NLnet with financial support from the European Commission's Next Generation Internet program, under the aegis of DG Communications Networks, Content and Technology under grant agreement No 825310.}
}

\author{\IEEEauthorblockN{Gabriel K. Gegenhuber\IEEEauthorrefmark{1}\IEEEauthorrefmark{2},
Wilfried Mayer\IEEEauthorrefmark{3}\IEEEauthorrefmark{4} and Edgar Weippl\IEEEauthorrefmark{5}} %
\IEEEauthorblockA{University of Vienna\IEEEauthorrefmark{2}, SBA Research\IEEEauthorrefmark{4}\\
Email: \IEEEauthorrefmark{1}gabriel.karl.gegenhuber@univie.ac.at,
\IEEEauthorrefmark{3}wmayer@sba-research.org,
\IEEEauthorrefmark{5}edgar.weippl@univie.ac.at}}

\maketitle

\def\toolname{\textsc{MobileAtlas}\xspace}

\begin{abstract}
\input{content/00_abstract}

\end{abstract}


\input{content/01_introduction}
\input{content/02_related_work}

\input{content/03_methodology}
\input{content/04_results}

\input{content/05_discussion}

\input{content/06_conclusion}

\bibliographystyle{IEEEtran}
\bibliography{IEEEabrv,bibliography}

\end{document}

%% file: content/00_abstract.tex
Zero-rating, the practice of not billing data traffic that belongs to certain applications, has become popular within the mobile ecosystem around the globe.
There is an ongoing debate whether mobile operators should be allowed to differentiate traffic or whether net neutrality regulations should prevent this.
Despite the importance of this issue, we know little about the technical aspects of zero-rating offers since the implementation is kept secret by mobile operators and therefore is opaque to end-users and regulatory agencies.

This work aims to independently audit classification practices used for zero-rating of four popular applications at seven different mobile operators in the EU.
We execute and evaluate more than 300 controlled experiments within domestic and internationally roamed environments and identify potentially problematic behavior at almost all investigated operators.
With this study, we hope to increase transparency around the current practices and inform future decisions and policies.

%% file: content/01_introduction.tex
\section{Introduction}
\label{sec:introduction}

Cellular networks have become a major access technology to the public Internet that can also be used across national borders.
In June 2017, the European Union abolished data roaming fees for the intra-EU/EEA area under the ``roam like at home'' doctrine. This regulation made roaming in foreign cellular networks feel and behave like at the home operator and led to a drastic increase in roaming traffic~\cite{Roaming2020}.
Mobile broadband connections do not only differ from landline data connections from a usage perspective but also diverge in terms of tariff models.
According to BEREC~\cite{ZeroRating2021}, a growing number of mobile network operators (MNOs) in the EU have introduced differential pricing offers (e.g., zero-rating), and some of them have already been taken to court~\cite{Curia2021} by regulators for disrespecting net neutrality principles (e.g., not applying zero-rating during intra-EU data roaming).
Obviously, correct zero-rating is crucial to consumer protection since misclassification can lead to illegitimately billed units for customers.
Possibly, many net neutrality violations remain undiscovered and proving an operator's misbehavior is not easily possible for end-users or regulatory agencies.

To address these issues, this work independently audits zero-rating practices of selected operators and gives valuable insights into the classification metrics that are currently used within the industry.

Our main contributions are:
\begin{itemize}
\item We propose a methodological approach to probe web endpoints for zero-rating.
\item We use this approach to evaluate zero-rating of four popular applications at seven operators from three countries.
\item We test the effect that intra-EU roaming has on zero-rating by executing our experiments during domestic and roaming usage scenarios in eight different countries.
\item We evaluate our results and give an overview of the used classification metrics and encountered classification errors.
\end{itemize}

The remainder of this paper is organized as follows.
Section~\ref{sec:related-work} gives an overview of related studies that are relevant to this work.
In Section~\ref{sec:methodology}, we describe our methodological approach and quickly introduce the testbed that was used to execute our experiments.
Section~\ref{sec:results} summarizes the results that were collected throughout this study.
Finally we discuss our results in Section~\ref{sec:discussion} and conclude with Section~\ref{sec:conclusion}.

%% file: content/02_related_work.tex
\section{Related Work}
\label{sec:related-work}
Aside from hot debates about political and regulatory decisions, net neutrality has also been an interesting research target from a technical perspective.
Net neutrality measurements usually aim to detect traffic differentiation in terms of
\begin{itemize}
  \item rate limiting (traffic shaping, traffic policing), %
  \item traffic blocking (censorship),
  \item traffic manipulation,
  \item economic differentiation (differential pricing, zero-rating).
\end{itemize}

\noindent\textbf{Identifying traffic differentiation.}
\textit{Glasnost}~\cite{dischinger2010glasnost} and \textit{NetPolice}~\cite{zhang2009detecting} were among the first tools that detect bandwidth throttling of specific protocols (e.g., BitTorrent).
While those tools were built for landline and desktop environments, \textit{BonaFide}~\cite{bashko2013bonafide} was released as a smartphone application. It replicates the capabilities of \textit{Glasnost} to the mobile world, minimizes data consumption, and supports a broader set of protocols (e.g., SIP and video streaming).
In contrast to prior work, \textit{Differentiation Detector}~\cite{kakhki2015identifying} and ~\textit{Wehe}\protect\footnote{\url{https://wehe.meddle.mobi}}~\cite{li2019large} do not target specific protocols, but moved to a application-centric design that mimics arbitrary protocols. By replaying pre-recorded application-generated traffic, they can treat it as a ``black box'' and do not need to provide specific implementation details of occurring protocols.

Although the mentioned tools were created to detect traffic differentiation in terms of rate-limiting, economic differentiation is usually built upon the same classifiers and therefore relies on similar metrics.
A study that investigates traffic classification in middleboxes~\cite{li2016classifiers} has shown that the used policies are often relatively simple and only match certain keywords within HTTP/HTTPS fields.
In contrast, our work does not only focus on keyword-based classifiers but also shows that IP-based classification is in widespread use by many operators.%

\noindent\textbf{Economic differentiation.}
Differential pricing practices like zero-rating or application-specific data quotas were rarely seen in the fixed-line landscape but have become common over the past years in the cellular field.
Although there is one case study investigating a zero-rating offer by T-Mobile (US) that targets video streaming~\cite{kakhki2016bingeon}, there is no work that compares current zero-rating practices across different operators or countries.

Our work aims to cover this research gap by comparing the used classification rules of popular applications at different operators in various European countries. Besides investigating domestic usage we also take the roaming context into account.

\noindent\textbf{Emerging technologies.}
The Internet and the used communication protocols are under constant change and evolution.
Previous work~\cite{czyz2016don, gbur2021quic} has shown that emerging technologies, such as IPv6 and QUIC (which is used for UDP-based communication at HTTP3), can cause problems with existing security policies or firewall configurations that are rather static.

Our work investigates whether current classification mechanisms also work with cutting-edge technology (i.e., IPv6 and HTTP3) that is already used by popular (zero-rated) applications in the wild.

%% file: content/03_methodology.tex
\section{Methodology}
\label{sec:methodology}
There are several parts to our methodology: analyzing the European cellular market and available zero-rating offers,
characterizing web endpoints that are used within zero-rated applications,
and using our testbed to determine which metrics are applied to classify the data traffic corresponding to a certain application.

\subsection{Market, Tariff, and Application Analysis}
To find out which countries and providers offer zero-rating tariffs, we conducted an EU-wide market analysis.
It has become increasingly popular that MNOs lease their wireless network infrastructure to Mobile Virtual Network Operators (MVNOs) that offer services to their customers but do not own any infrastructure. Compared to an MNO, becoming an MVNO is relatively easy and requires less financial effort. Thereby, many countries have got a vast amount of operators (e.g., Austria currently has about 40 MVNOs, despite being a relatively small country). However, due to well-established MNOs and the high fluctuation of MVNOs, the latter usually play a minor role in terms of actual market penetration. To limit the effort but accordingly respect the market situation, we limited our market analysis to bare-metal consumer-grade MNOs in every country.
After identifying the relevant players, we looked at the available tariffs to find out whether they offer differential pricing or zero-rating programs.
For this step, our primary source of information was a provider's website.
The language barrier at foreign countries and complex tariff structures (e.g., prepaid vs. postpaid, minimum contract duration, packages that are only available in certain tariffs) often made it cumbersome to get the required information.

According to our market analysis that was conducted in May 2021, operators in 24 EU countries (ca. 89\%) have implemented differential pricing or zero-rating offers. 
SIM card registration is currently required in 14 of 27 EU countries (ca. 52\%)~\cite{SimReg}.

We selected three countries (Austria, Croatia, and Romania) that offered a good coverage of zero-rating tariffs and were available in our testbed environment and then acquired SIM cards of the relevant MNOs. In Austria, all three MNOs implement zero-rating, while in Croatia and Romania, two out of three providers offer relevant tariffs. Table~\ref{tab:selected-tariffs} shows an overview of the applications that are included in the analyzed tariffs at each operator.

\begin{table}[bp] %
\caption{Overview of the selected tariffs and zero-rated applications}
\begin{center}
\begin{tabularx}{\columnwidth}{l X}
\toprule
Operator & Included Applications (Zero-Rating)                                                         \\ \midrule
AT-1      & WhatsApp, Snapchat, Messenger, Viber                                                       \\
AT-2      & WhatsApp, Snapchat, Facebook, Instagram, TikTok, Twitter                      \\
AT-3      & WhatsApp, Messenger, FM4, Ö3, Ö1, Energy, Superfly, Antenne, 886, Kronehit, Radio Arabella \\
HR-1      & WhatsApp, Snapchat, Messenger, Facebook                                              \\
HR-2      & WhatsApp, Snapchat, Messenger, Facebook, Instagram, TikTok                                 \\
RO-1      & WhatsApp, Facebook, TikTok                                                                 \\
RO-2      & WhatsApp, Message+                                                                         \\ \bottomrule
\end{tabularx}
\label{tab:selected-tariffs}
\end{center}
\end{table}

For our measurements, we selected the four topmost included applications: WhatsApp (seven operators), Snapchat (four operators), Messenger (four operators), and Facebook (four operators).

Although our target applications use a variety of communication protocols (e.g., XMPP, RTP, or MQTT) they usually rely on Content Delivery Networks (CDNs) that distribute the actual content via web (i.e., HTTPS). Legacy protocols often use TLS connections to encrypt the data (e.g., SRTP, or MQTTS).
Since those web endpoints are responsible for a substantial part of the data traffic, this needs to be zero-rated by the operator. 

To determine relevant web endpoints that are used within the applications of interest, we obtained exemplary traffic dumps for each application. We ensured to use the most recent Android applications from the AppStore and recorded the data traffic that occurred within five minutes of application usage via \textit{PCAPdroid}\protect\footnote{\url{https://github.com/emanuele-f/PCAPdroid}}.
Since most operators explicitly exclude external and advertisement content from zero-rating in their Terms of Service (ToS), we ensure to stay within the original application during our recording sessions. The same applies to voice- and video calls (e.g., via WhatsApp), which often are exempted from the zero-rating offer.
To check whether an application or web endpoint supports different technology stacks (e.g., IPv4/IPv6, or HTTPS/HTTP3) we record multiple traffic dumps in varying testing environments (e.g., IPv4 only, IPv6 only, or Dual Stack).

By analyzing our traffic dumps, we chose one web endpoint for each application that causes a substantial part of data traffic and, furthermore, supports a broad set of communication protocols.
Table~\ref{tab:selected-applications} shows which endpoints were selected for the probed applications.
All selected web endpoints support both IPv4 and IPv6, as well as HTTP, HTTPS, and HTTP3.
Notably, the endpoints which are used by Messenger and Facebook overlap to a high degree.
Presumably, one application cannot be separately zero-rated without also triggering classification of traffic that belongs to the other application.

\begin{table}[htbp]%
\caption{Web endpoints and resources that were used to test for classification}
\begin{center}
\newcommand\tFa{$^{\mathrm{a}}$}
\newcommand\tFb{$^{\mathrm{b}}$}
\newcommand\tFc{$^{\mathrm{c}}$}
\begin{tabular}{@{}lll@{}}
\toprule
Application & Endpoint              & Used Resource  \\ \midrule
WhatsApp    & static.whatsapp.net   & Logo\tFa       \\ %
Snapchat    & app.snapchat.com      & User Avatar\tFb\\ %
Messenger   & scontent.xx.fbcdn.net & Favicon\tFc    \\
Facebook    & scontent.xx.fbcdn.net & Favicon\tFc    \\ \bottomrule
\multicolumn{3}{l}{\tFa~\url{static.whatsapp.net/rsrc.php/v3/yP/r/rYZqPCBaG70.png}}\\
\multicolumn{3}{l}{\tFb~\url{app.snapchat.com/web/deeplink/snapcode}}\\
\multicolumn{3}{l}{\tFc~\url{scontent.xx.fbcdn.net/favicon.ico}}
\end{tabular}
\label{tab:selected-applications}
\end{center}
\end{table}

\subsection{Testbed}
As described in Section~\ref{sec:related-work}, traffic differentiation measurements that aim to detect rate limiting of certain applications are often crowd-sourced and executed on volunteers' smartphones.
This does not work for detecting economic differentiation since we do not have insights into the data units that are billed in tariffs of foreign entities. Furthermore, unwanted background activities (e.g., traffic that is caused by the smartphone user) might distort measurement results.
Therefore, we need controlled experiments on dedicated SIM cards where only explicit data traffic is transmitted.

\begin{figure}[htpb]
    \centering
	\includegraphics[width=\columnwidth]{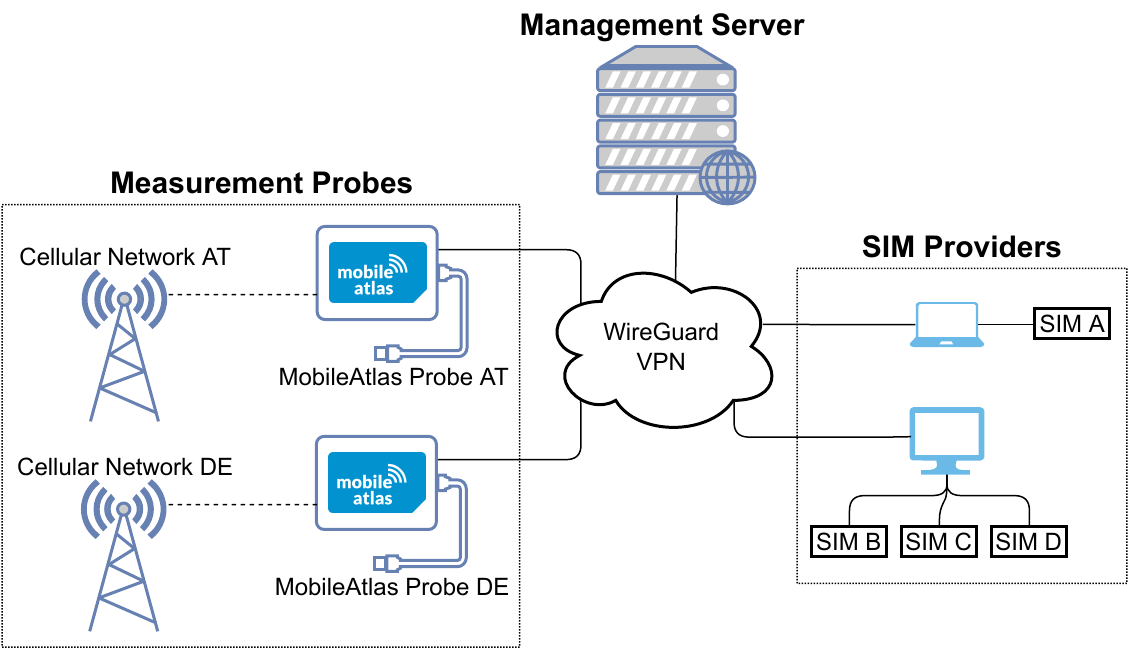}
    \caption{Architecture and components of the \toolname measurement platform}
    \label{fig:architecture}
\end{figure}
To execute measurements within a domestic and various roaming environments we used the \toolname\protect\footnote{\url{https://www.mobileatlas.eu}} measurement platform.
\toolname geographically decouples SIM card and modem by tunneling the SIM card's protocol over the Internet and emulating its signal on the LTE modem.
This boosts the scalability and flexibility of international measurements in the cellular field because it allows testing roaming effects on a large number of operators without physically moving any hardware between different countries.
The platform is currently deployed in eight European countries: Austria, Belgium, Croatia, Finland, Germany, Romania, Slovakia, and Slovenia. As Fig.~\ref{fig:architecture} shows, the framework can be structured into three main components: SIM providers that allow sharing SIM card access, measurement probes that act as a local breakout to the cellular network, and a management server that connects the prior two components and acts as command and control unit for the measurement probes.

To execute a measurement, any SIM card that is attached to our system via a reader device (e.g., a PC/SC reader) can be virtually connected with a measurement probe in the desired target country. \toolname provides various interfaces to interact with the measurement probe's modem and isolates the cellular network connection from any unwanted noise sources. For billing-related measurements, it offers a template for credit checking that can leverage different modem capabilities (e.g., sending SMS messages or USSD-codes and a separate network gateway to access the customer zone) to retrieve the current data quota of a SIM card.

Since there is no standardized interface that allows retrieving the available credit information across different operators, we needed to provide a specific implementation for each tariff tested in this study.
Most operators implement credit retrieval through one or more of the following ways: SMS message, USSD code, voice call, website (customer area), or mobile app.
\begin{table}[bp]%
\caption{Operator-wise overview of used credit checking method}
\begin{center}
\begin{tabular}{@{}ll@{}}
\toprule
Operator & Credit Checking Method \\ \midrule
AT-1     & SMS (domestic), APP (roaming)                    \\
AT-2     & APP                    \\
AT-3     & APP                    \\
HR-1     & APP                    \\
HR-2     & APP                    \\
RO-1     & APP (+ SMS for OTP Login) \\
RO-2     & USSD                   \\ \bottomrule
\end{tabular}
\label{tab:used-credit-checking}
\end{center}
\end{table}
We analyzed the available methods for all operators and usually chose the approach that provided the most verbose billing information.
The granularity in which the current data quota is reported differs heavily from fine-grained reporting in byte precision to coarse gigabyte estimates between each operator and retrieval method.
Table~\ref{tab:used-credit-checking} shows that we use the mobile app approach in most cases but opted for the SMS and USSD approach for one operator each.
In one case (RO-1), the password-based login form required to solve a captcha, which forced us to use an alternative login page that uses a one-time password (OTP) that is sent via SMS.
To find the appropriate API endpoints that are used within the apps to retrieve the credit from the network, we had to reverse engineer each provider's mobile app.
Depending on the application, we used various approaches and tools (e.g., static/dynamic analysis, JADX, Frida, and Burp Suite) to determine how a provider's app retrieves a customer's credit.

\subsection{Measurement Implementation}
For tests on differential pricing, we need to know whether specific traffic is deducted from the available credit units or funds.
To cope with different update latency of consumed units and to enable running multiple payloads without in-between waiting for the billing records to update, we use binary exponents, i.e., every payload uses traffic amounts selected from $\mathit{base unit} \times 2^\mathit{index}$.
For example, when the first payload causes one megabyte of traffic, the second has two megabytes, the third four megabytes, and so on.
When the final traffic billing arrives (which in our case is a control payload that is always billed), we can unambiguously deduct which payloads were counted towards our tariff's quota.

To reveal potential metrics that are used for classification, we designed three different experiments: at first, we verify that an endpoint is actually zero-rated by an operator, secondly, we check for IP-based classification, and lastly, we check for hostname-based classification.

\subsubsection{Verify Zero-Rating}
\begin{figure}[htpb]
    \centering
	\includegraphics[height=4.6cm]{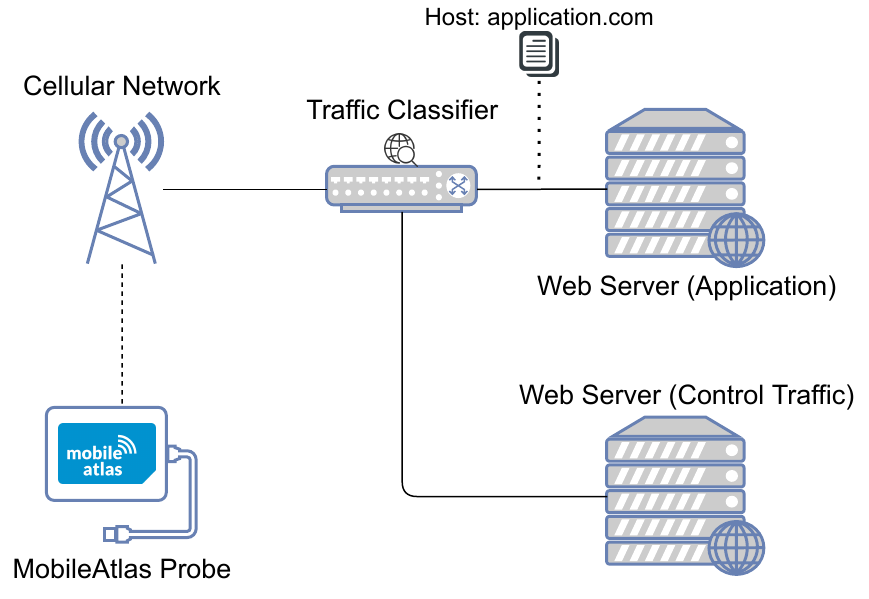}
    \caption{Involved actors and traffic flow when verifying zero-rated data traffic}
    \label{fig:verify-zero-rating}
\end{figure}

In the first experiment, we cause regular data traffic to our examined web endpoint.
Fig.~\ref{fig:verify-zero-rating} describes the involved actors and the traffic flow that occurs during verification for an endpoint with the hostname \textit{application.com}.
We always query a static resource (e.g., the favicon) that actually is present at the target web server to increase the response size and speed up our experiments. Alternatively, we can request the web server's root or index file, which does not require any additional knowledge about the probed web application but is usually slower because of possible \texttt{HTTP 404}-responses.
We repeatedly send out web queries until the caused data traffic reaches our target size (e.g., one megabyte).
To minimize unwanted background noise, we ensure that DNS queries are only issued once and cached locally for subsequent requests.
Finally, the test generates control traffic to a third party that is not part of any zero-rating program and, therefore, normally billed.
As previously described, the test terminates as soon as the control traffic is recognized.
To execute this experiment for multiple protocols (i.e., HTTP, HTTPS, and HTTP3), we use the payload multiplexing technique that is described at the beginning of this section.

\subsubsection{Detect IP-based Classification}
\begin{figure}[pb]
    \centering
	\includegraphics[height=4.6cm]{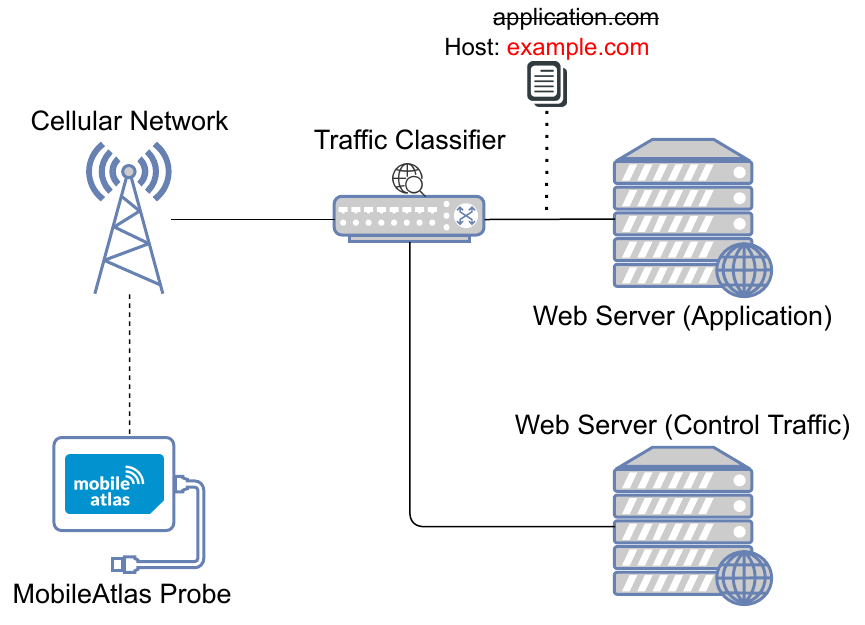}
    \caption{Actors and traffic flow when checking for IP-based classification}
    \label{fig:detect-ip-based-classification}
\end{figure}
This experiment aims to expose IP-based classification rules.
As Fig.~\ref{fig:detect-ip-based-classification} shows, the involved actors have not changed from the previous step.
We connect to the application's web server but replace the hostname that is sent in an HTTP request or within the initial TLS handshake (i.e., the SNI header for HTTPS or HTTP3).
Although the packets are sent to the real application's web server (i.e., its actual IP address), they do not contain the real endpoint's hostname because it was exchanged with a dummy value (i.e., \textit{example.com}).

When our test traffic is nevertheless zero-rated by an operator, the used classification is presumably based on IP addresses. In case of billed test traffic, we suppose that hostname-based classification rules were used.

\subsubsection{Detect Hostname-based Classification}
\begin{figure}[htpb]
    \centering
	\includegraphics[height=4.6cm]{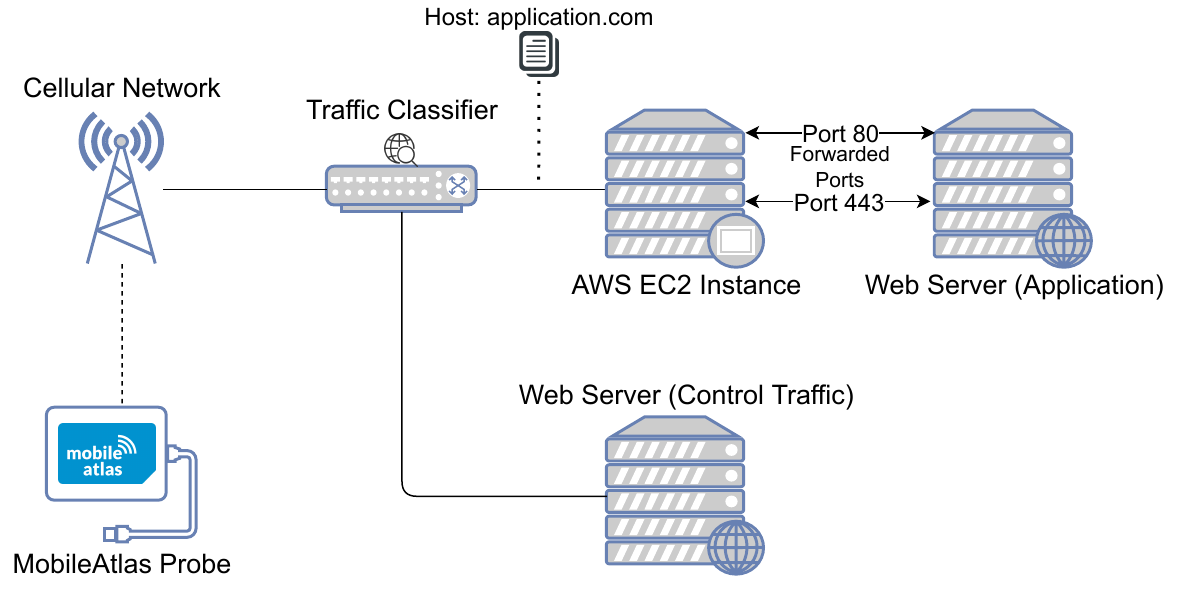}
    \caption{Involved actors and traffic flow when checking for hostname-based classification}
    \label{fig:detect-host-based-classification}
\end{figure}
The first step at this experiment is to retrieve the IP address of the server that holds the examined web resource.
Secondly, an Amazon EC2-instance is launched automatically and forwards the corresponding ports for the protocols that should be tested (e.g., \texttt{TCP80} and \texttt{TCP443} for HTTP and HTTPS and \texttt{UDP443} for HTTP3).
Thus, when a TCP connection to the freshly spawned EC2 server is initiated on port 80, the connection is forwarded to the original web server. Thereby, the same content is served, although the data packet that is processed by the provider is headed to a different IP address.
When executing the payload for a certain protocol, the measurement environment pins the hostname of the original web resource to the IP address of the EC2 instance. Therefore, the measurement is conducted against a third-party IP address (i.e., against the EC2 server).
Fig.~\ref{fig:detect-host-based-classification} gives an overview of the involved actors and the traffic flow during this experiment. When the data packets are passing the classifier, the hostname within the packets matches the one from the application. However, the IP address of the packets does not match the address of the application's web server because the packets are headed to the EC2 instance. Yet, the content of the data packets is equal to the previous test because the EC2 instance simply acts as a proxy to the actual application's web server.

When our test traffic is zero-rated by an operator, we imply that hostname-based classification was used.

\subsection{Ethical Considerations}
Ethical considerations are vital to the field of measurements, especially with active measurements conducted in live systems.
We tried to reflect normal user behavior whenever possible, e.g., by introducing a minimum waiting time between switching to another country with a SIM card. Furthermore, we took care not to overstress any mobile operator's infrastructure we were interacting with (e.g., by rate-limiting credit retrieval).
Because our measurements might still stay in conflict with an operator's ToS and possibly lead to blocked SIM cards, we ensured only to use SIM cards that were exclusively purchased for measurements in this paper.
We conduct our measurements to get a better understanding of current traffic differentiation mechanisms.
This can be of advantage for a variety of stakeholders, including but not limited to: customers, content providers, 
regulatory authorities, policy makers, and MNOs.

%% file: content/04_results.tex
\section{Experiment Results}
\label{sec:results}
We executed the experiments that are described in Section~\ref{sec:methodology} for all selected operators and applications and tested classification metrics for the chosen endpoints using HTTP, HTTPS, and HTTP3.

Table~\ref{tab:summarized-results} summarizes the results of our experiments that were executed within two measurement periods (September 2021 and May 2022).
Some operators combine IP- and hostname-based classification for a single endpoint.
In such cases, the traffic is zero-rated when at least one of the two rules applies.

\begin{table}[tbp]
\caption{Used classification metrics at the tested operators and applications}
\begin{center}
\newcommand\tBilled{\$}
\newcommand\tNoAvail{$\times$}
\newcommand\tFa{$^{\mathrm{a}}$}
\newcommand\tFb{$^{\mathrm{b}}$}
\newcommand\tFc{$^{\mathrm{c}}$}
\begin{tabular}{@{}lcccc@{}}
\toprule
Operator & Roaming   & WhatsApp      & Snapchat     & Messenger/Facebook \\ \midrule
AT-1     & Yes       & IP            & IP, Host     & \tBilled           \\
AT-2     & Yes       & IP            & IP\tFa       & IP                 \\
AT-3     & Yes       & IP            & \tNoAvail    & \tBilled           \\
HR-1     & No        & IP            & Host         & IP                 \\
HR-2     & Yes       & IP            & IP, Host\tFb & IP                 \\
RO-1     & No        & IP, Host\tFb  & \tNoAvail    & IP, Host\tFb       \\
RO-2     & \tNoAvail & IP\tFc        & \tNoAvail    & \tNoAvail          \\ \bottomrule
\end{tabular}\\
\rule{0in}{1.5em}
\tBilled~traffic fully billed.\hspace{1ex}
\tNoAvail~not part of zero-rating tariff.\hspace{1ex}\\
\tFa~IPv4 only.\hspace{1ex}
\tFb~HTTPS only.
\tFc~TCP only.\hspace{1ex}\\
\label{tab:summarized-results}
\end{center}
\end{table}

\noindent\textbf{General misclassification.}
Our measurements indicate that several operators use wrongfully configured classification and therefore bill traffic to their customers that should actually be zero-rated.
For example, two Austrian operators (AT-1 and AT-3) do not zero-rate any traffic that goes to our selected web endpoint of the Messenger app.
In both cases, we verified that there is indeed a classification problem in practice by plugging the SIM card into a smartphone, using the Messenger app, and capturing the caused data traffic. We took snapshots of the available data quota before and after the test and could verify that only minor parts of the occurred data traffic were zero-rated (e.g., only 0.47 out of 17.39 megabytes in a randomly selected test sample).

\noindent\textbf{IPv6 and HTTP3.}
At the two operators (AT-1 and AT-2) that already deployed IPv6 (DualStack) to their customers, we also tested whether accessing an endpoint via IPv6 instead of IPv4 influences traffic classification.
The operator AT-2 does only zero-rate traffic that goes to the IPv4 address of the Snapchat endpoint but fully bills data packets that go to the corresponding IPv6 address.

For HTTP3 we observed similar behaviour at multiple providers.
The provider HR-1 relies on hostname-based classification for the Snapchat endpoint.
During our first measurement period in September 2021, HTTP3 traffic to Snapchat was wrongfully billed.
However, in May 2022, the same traffic was correctly zero-rated (i.e., the operator updated the responsible classification rules).

Similarly, AT-1 upgraded its hostname-based classification metrics to work with HTTP3 between our two measurement periods.
In contrast to HR-1, the traffic at AT-1 was nevertheless classified correctly with the old metrics because a combined classification approach was used, and the additional IP-address rule ensured the correct classification of HTTP3. %

Lastly, the IP-based classification for the selected WhatsApp endpoint does not work with HTTP3 at the operator RO-2.

\noindent\textbf{Roaming.}
For all tested tariffs that include EU-wide roaming in their tariffs (i.e., all, except RO-2), we checked whether zero-rating is active when connecting from a foreign country.
To examine whether there is any difference in classification between local and roamed connections, we executed our classification experiments in all eight countries that are available in our testbed at each provider.
To choose the appropriate roaming partner, we used automatic network selection.
We did not observe any cases of local-breakout, since all operators used home-routing to terminate their roamed data connections.

We observed that four operators (AT-1, AT-2, AT-3, and HR-2) also applied zero-rating during roaming, while two operators (HR-1 and RO-1) had zero-rating disabled and fully billed the occurring traffic.
We did not notice any differences between different visited countries (e.g., our results in Germany were identical to the results measured in Finland).

%% file: content/05_discussion.tex
\section{Discussion and Limitations}
\label{sec:discussion}

The results of our experiments show that several operators:
\begin{itemize}
    \item wrongfully bill a substantial amount of traffic from zero-rated applications.
    \item do not correctly apply zero-rating when newer technology stacks (i.e., IPv6 or HTTP3) are used.
    \item turn off zero-rating during intra-EU roaming, although this has already been impeached by regulatory agencies in several countries.
\end{itemize}

Overall, we were surprised by the huge amount of wrongfully billed traffic.
Although we noticed that several operators adapted their rules during our two measurement periods to support HTTP3 connections at Snapchat, it is already used by the application for at least one year~\cite{Snapchat2021}.
From a customer's perspective, paying for traffic that is already part of the purchased tariff is not acceptable.

We believe that operators should be more transparent and make the technical implementation of zero-rating offers publicly available. This would enable Internet activists and regulatory authorities to find classification errors more quickly and prevent wrongfully billed traffic.

Our methodology is built to detect simple classification metrics (i.e., IP- or hostname-based classification).
Theoretically, an operator could also deploy more complex algorithms (e.g., traffic fingerprinting, machine learning~\cite{aceto2019mobile}) to detect data packets that belong to a certain application. Furthermore, our approach could trigger anomaly detection when repeatedly requesting one single web resource.
However, the amount of traffic that was caused by our experiments was considerably small (e.g., one megabyte), and we did not observe anything that would hint at being flagged by an operator.
To verify the lack of zero-rating for a certain application, we usually ran additional tests by manually plugging the SIM card into a smartphone and observing the billed units when the probed application is in actual use.

Although our used testbed allows rather flexible and mostly automated measurements, it still requires effort to run and evaluate these experiments, limiting our tests' coverage.
Since one application usually communicates with a plethora of web endpoints, it remains to future work to improve test automatability and run more verbose experiments.

%% file: content/06_conclusion.tex
\section{Conclusion}
\label{sec:conclusion}
This paper provides a technical analysis of differential pricing practices at European MNOs.
Our analysis shows that operators currently use both IP- and hostname-based classification.
Moreover, we show that several operators do not correctly classify the traffic of certain applications or when particular protocols are used.
Furthermore, we demonstrate that some providers do not apply zero-rating in roaming usage scenarios.

To encourage other researchers to look into the topic and improve the available tools for controlled and international cellular measurements, we've open-sourced the implementation of our experiments alongside our \toolname testbed and will publish all collected measurement artifacts that were used within this study upon publication.
We hope that this study will help to inform future decisions and policies around net neutrality.